\documentclass[12pt,twoside]{article}

\evensidemargin=\oddsidemargin
\def\Title#1#2#3{%
    \baselineskip=18pt
    \begin{center}
          {\large\bf\uppercase{#1} \\ }
          \bigskip\bigskip
          {#2} \\
          {#3} \\
    \end{center}}
\long\def\Abstract#1{%
         \bigskip
         \parbox{0.93\textwidth}{%
                 \begin{center}
                       {\bf Abstract} \\
                 \end{center}
                 \medskip{\baselineskip=14pt #1}
                 \vss}
         \bigskip}

\makeatletter
\renewcommand{\section}%
 {\@startsection{section}{1}{0pt}%
  {-3.25ex plus -1ex minus -.2ex}{1.5ex plus .2ex}%
  {\vspace*{5mm}\raggedright\large\bf }}
\renewcommand{\thesection}{\arabic{section}.}
\@addtoreset{equation}{section}
\renewcommand{\@eqnnum}{(\thesection\theequation)}
\renewcommand{\p@equation}{\thesection}
\makeatother

\begin{document}

\vspace*{1cm}

\Title{COULD GAUGE GRAVITATIONAL DEGREES OF FREEDOM\\
PLAY THE ROLE OF ENVIRONMENT\\
IN ``EXTENDED PHASE SPACE'' VERSION\\
OF QUANTUM GEOMETRODYNAMICS}%
{T. P. Shestakova}%
{Department of Theoretical and Computational Physics, Rostov State University, \\
Sorge Str. 5, Rostov-on-Don 344090, Russia \\
E-mail: {\tt shestakova@phys.rsu.ru}}

\Abstract{In the context of the recently proposed formulation of quantum
geometrodynamics in extended phase space we discuss the problem how
the behavior of the Universe, initially managed by quantum laws,
has become classical. In this version of quantum geometrodynamics
we quantize gauge gravitational degrees of freedom on an equal
basis with physical degrees of freedom. As a consequence of this
approach, a wave function of the Universe depends not only on
physical fields but also on gauge degrees of freedom. From this
viewpoint, one should regard the physical Universe as a subsystem
whose properties are formed in interaction with the subsystem of
gauge degrees of freedom. We argue that the subsystem of gauge
degrees of freedom may play the role of environment, which, being
taken into account, causes the density matrix to be diagonal. We
show that under physically reasonable fixing of gauge condition the
density matrix describing the physical subsystem of the Universe
may have a Gaussian peak in some variable, but it could take the
Gaussian form only within a spacetime region where a certain gauge
condition is imposed. If spacetime manifold consists of regions
covered by different coordinate charts the Universe cannot behave
in a classical manner nearby borders of these regions. Moreover, in
this case the Universe could not stay in the same quantum state,
but its state would change in some irreversible way.}

\section{Introduction}
Quantum geometrodynamics claims to give a description of quantum
stage of the Universe evolution, including an explanation of the
fact that the behavior of the Universe, initially managed by
quantum laws, has become classical. According to Halliwell
\cite{Halliwell}, two requirements must be satisfied for a system
to be regarded as classical. The first requirement is that its
evolution should be described by classical laws in a very good
approximation, and the second requirement involves the notion of
decoherence -- a transition from a pure state to some mixed state
described by diagonal density matrix. There has been a number of
works where the notion of decoherence was discussed in the context
of quantum cosmology based on the Wheeler -- DeWitt quantum
geometrodynamics. As well known, the destruction of the
off-diagonal terms in the density matrix cannot be regarded in the
limits of unitary evolution. A possible way to solve this problem
is to consider this destruction as a result of interaction with
some environment \cite{Zurek, UZ}. The application of this idea to
quantum cosmology implies splitting the Universe into two
subsystems, one of which is a system under investigation and the
other plays the role of environment. It has been suggested to
consider some modes of scalar, gravitational and other fields as an
environment (see, for example, \cite{Halliwell, FM}). However,
there is no natural way to split the Universe into two subsystems.

The aim of the present work is to discuss these questions in the
limits of recently proposed formulation of quantum geometrodynamics
in extended phase space \cite{SSV1} -- \cite{Shest3}. In this
version of quantum geometrodynamics we quantize gauge gravitational
degrees of freedom on an equal basis with physical degrees of
freedom. The motivation for it was that it is impossible to
separate gauge, or "non-physical" degrees of freedom from physical
ones if the system under consideration does not possess asymptotic
states, and it is indeed the case for a closed universe as well as
for a universe with rather nontrivial topology. As a consequence of
this approach, a wave function of the Universe depends not only on
physical fields but also on gauge degrees of freedom. From this
viewpoint one should regard the physical Universe as a subsystem
whose properties are formed in interaction with the subsystem of
gauge degrees of freedom.

The plan of the work is as follows. We remind basic equations of
the ``extended phase space'' version of quantum geometrodynamics,
consider a general structure of a wave function of the Universe and
construct a density matrix describing the physical subsystem of the
Universe. We then show that under physically reasonable fixing of
gauge condition the density matrix may have a Gaussian peak in some
variable, say, in a scale factor, while other degrees of freedom
(e. g., gravitational waves and matter fields) should be treated
quantum mechanically. The important point in this consideration is
that the density matrix takes the Gaussian form only within a
spacetime region where a certain gauge condition is imposed. In
simple cosmological models one can introduce a single gauge
condition in the whole spacetime, and the behavior of the Universe
can be regarded as classical almost over the whole history of the
Universe. Meanwhile, in the case of nontrivial topology spacetime
manifold may consist of regions covered by different coordinate
charts, so that one should impose different gauge conditions in
these regions. Then, the Universe cannot behave in a classical
manner nearby borders of these regions. We shall argue that in this
case the Universe could not stay in the same quantum state; as a
consequence of interaction with the subsystem of gauge degrees of
freedom its state would change in some irreversible way.

\section{The ``extended phase space'' version of quantum geometrodynamics: basic formulas}
To investigate a system without asymptotic states we make use of
the path integral approach \cite{SSV3, SSV4}. It is easy to
illustrate the crux of the matter for a simple minisuperspace model
with a gauged action
\begin{equation}
\label{action}
S=\!\int\!dt\,\biggl\{
  \displaystyle\frac12 v(\mu, Q)\gamma_{ab}\dot{Q}^a\dot{Q}^b
  -\frac1{v(\mu, Q)}U(Q)
  +\pi_0\left(\dot\mu-f_{,a}\dot{Q}^a\right)
  -i w(\mu, Q)\dot{\bar\theta}\dot\theta\biggr\}.
\end{equation}
Here $Q=\{Q^a\}$ stands for physical variables such as a scale
factor or gravitational-wave degrees of freedom and material
fields, and we use an arbitrary parametrization of a gauge variable
$\mu$ determined by the function $v(\mu, Q)$. For example, in the
case of isotropic universe or the Bianchi IX model $\mu$ is bound
to the scale factor $a$ and the lapse function $N$ by the relation
\begin{equation}
\label{paramet}
\displaystyle\frac{a^3}{N}=v(\mu, Q).
\end{equation}
$\theta,\,\bar\theta$ are
the Faddeev -- Popov ghosts after replacement
$\bar\theta\to -i\bar\theta$. Further,
\begin{equation}
\label{w_def}
w(\mu, Q)=\frac{v(\mu, Q)}{v_{,\mu}};\quad
v_{,\mu}\stackrel{def}{=}\frac{\partial v}{\partial\mu}.
\end{equation}
We confine attention to the special class of gauges not depending
on time
\begin{equation}
\label{frame_A}
\mu=f(Q)+k;\quad
k={\rm const},
\end{equation}
which can be presented in a differential form,
\begin{equation}
\label{diff_form}
\dot{\mu}=f_{,a}\dot{Q}^a,\quad
f_{,a}\stackrel{def}{=}\frac{\partial f}{\partial Q^a}.
\end{equation}

The Schr\"odinger equation for this model reads
\begin{equation}
\label{SE1}
i\,\frac{\partial\Psi(\mu,Q,\theta,\bar\theta;\,t)}{\partial t}
 =H\Psi(\mu,\,Q,\,\theta,\,\bar\theta;\,t),
\end{equation}
where
\begin{equation}
\label{H}
H=-\frac i w\frac{\partial}{\partial\theta}
   \frac{\partial}{\partial\bar\theta}
  -\frac1{2M}\frac{\partial}{\partial Q^{\alpha}}MG^{\alpha\beta}
   \frac{\partial}{\partial Q^{\beta}}
  +\frac1v(U-V);
\end{equation}
$M$ is the measure in the path integral,
\begin{equation}
\label{M}
M(\mu, Q)=v^{\frac K2}(\mu, Q)w^{-1}(\mu, Q);
\end{equation}
\begin{equation}
\label{Galpha_beta}
G^{\alpha\beta}=\frac1{v(\mu, Q)}\left(
 \begin{array}{cc}
  f_{,a}f^{,a}&f^{,a}\\
  f^{,a}&\gamma^{ab}
 \end{array}
 \right);\quad
\alpha,\beta=(0,a);\quad
Q^0=\mu,
\end{equation}
$K$ is a number of physical degrees of freedom; the wave function
is defined on extended configurational space with the coordinates
$\mu,\,Q,\,\theta,\,\bar\theta$. $V$ is a quantum correction to the
potential $U$, that depends on the chosen parametrization
(\ref{paramet}) and gauge (\ref{frame_A}):
\begin{eqnarray}
V&=&\frac5{12w^2}\left(w^2_{,\mu}f_{,a}f^{,a}+2w_{,\mu}f_{,a}w^{,a}
    +w_{,a}w^{,a}\right)
   +\frac1{3w}\left(w_{,\mu,\mu}f_{,a}f^{,a}+2w_{,\mu,a}f^{,a}
    +w_{,\mu}f_{,a}^{,a}+w_{,a}^{,a}\right)+\nonumber\\
&+&\frac{K-2}{6vw}\left(v_{,\mu}w_{,\mu}f_{,a}f^{,a}+v_{,\mu}f_{,a}w^{,a}
    +w_{,\mu}f_{,a}v^{,a}+v_{,a}w^{,a}\right)-\nonumber\\
&-&\frac{K^2-7K+6}{24v^2}\left(v^2_{,\mu}f_{,a}f^{,a}+2v_{,\mu}f_{,a}v^{,a}
    +v_{,a}v^{,a}\right)+\nonumber\\
\label{V}
&+&\frac{1-K}{6v}\left(v_{,\mu,\mu}f_{,a}f^{,a}+2v_{,\mu,a}f^{,a}
    +v_{,\mu}f_{,a}^{,a}+v_{,a}^{,a}\right).
\end{eqnarray}

The Schr\"odinger equation (\ref{SE1}) -- (\ref{V}) is derived from
a path integral with the effective action (\ref{action}) without
asymptotic boundary conditions by the standard well-definite
Feynman procedure, thus it is {\it a direct mathematical
consequence} of the path integral. Once we agreed that imposing
asymptotic boundary conditions is not correct in the case of a
closed universe, we {\it are doomed} to come to a gauge-dependent
description of the Universe.

\section{The general solution to the Schr\"odinger equation and the density matrix}
The general solution to the Schr\"odinger equation (\ref{SE1}) has
the following structure \cite{SSV4}:
\begin{equation}
\label{GS-A}
\Psi(\mu,\,Q,\,\theta,\,\bar\theta;\,t)
 =\int\Psi_k(Q,\,t)\,\delta(\mu-f(Q)-k)\,(\bar\theta+i\theta)\,dk.
\end{equation}
The dependence of the wave function (\ref{GS-A}) on ghosts is
determined by the demand of norm positivity.

Note that the general solution (\ref{GS-A}) is a superposition of
eigenstates of a gauge operator,
\begin{equation}
\label{k-vector}
\left(\mu-f(Q)\right)|k\rangle=k\,|k\rangle;\quad
|k\rangle=\delta\left(\mu-f(Q)-k\right).
\end{equation}
It can be interpreted in the spirit of Everett's ``relative state''
formulation. In fact, each element of the superposition
(\ref{GS-A}) describe a state in which the only gauge degree of
freedom $\mu$ is definite, so that time scale is determined by
processes in the physical subsystem through functions
$v(\mu,\,Q),\,f(Q)$ (see (\ref{paramet}), (\ref{frame_A})), while
$k$ being determined by initial clock setting. Indeed, according to
(\ref{frame_A}), the parameter $k$ gives an initial condition for
the variable $\mu$. The function $\Psi_k(Q,\,t)$ describes a state
of the physical subsystem for a reference frame fixed by the
condition (\ref{frame_A}). It is a solution to the equation
\begin{equation}
\label{phys.SE}
i\,\frac{\partial\Psi_k(Q;\,t)}{\partial t}
 =H_{(phys)}\Psi_k(Q;\,t),
\end{equation}
\begin{equation}
\label{phys.H-A}
H_{(phys)}=\left.\left[-\frac1{2M}\frac{\partial}{\partial Q^a}
  \frac1v M\gamma^{ab}\frac{\partial}{\partial Q^b}
 +\frac1v (U-V)\right]\right|_{\mu=f(Q)+k}.
\end{equation}

In general, one can seek a solution to Eq.(\ref{phys.SE}) in the
form of superposition of stationary state eigenfunctions:
\begin{equation}
\label{stat.WF}
\Psi_k(Q,\,t)=\sum_n c_n\Psi_{kn}(Q)\exp(-iE_n t);
\end{equation}
\begin{equation}
\label{stat.states}
H_{(phys)}\Psi_{kn}(Q)=E_n\Psi_{kn}(Q).
\end{equation}
The eigenvalue $E$ corresponds to a new integral of motion that
emerges in the proposed formulation as a result of fixing a gauge
condition and characterizes the gauge subsystem (see below
Eq.(\ref{eigenvalue})).

In this paper we shall be interested under what conditions the
behavior of the Universe can be regarded as classical. As well
known, one of necessary requirements is that a wave function could
be represented in the WKB form: $\Psi_{kn}(Q)=C(Q)\exp[iS(Q)]$.
However, in our formulation there is an additional requirement. In
the classical limit the Universe is described by gauge-invariant
Einstein equations, so that all vestiges of gauge fixing should be
eliminated. In particular, $E$ must take the zero eigenvalue. Thus,
in the classical limit the Universe appears to be in the unique
eigenstate with $E=0$.

Strictly speaking, we need some mechanism of the "reduction" of the
wave function (\ref{stat.WF}) to the state with $E=0$. We suppose
that such a mechanism involves a specific interaction between gauge
and physical subsystems, but we have not been able to give
explanation of the mechanism. In this paper we shall assume that,
in any region where the quasiclassical approximation exists, the
Universe is described by a quasiclassical wave function of the
special state with $E=0$, $\Psi_k(Q,\,t)=\Psi_{kn}(Q)$, and
concentrate on the density matrix of the physical subsystem.

The normalization condition for the wave function (\ref{GS-A})
reads
$$
\int\Psi^*(\mu,\,Q,\,\theta,\,\bar\theta;\,t)\,
 \Psi(\mu,\,Q,\,\theta,\,\bar\theta;\,t)\,M(\mu,\,Q)\,
 d\mu\,d\theta\,d\bar\theta\,\prod_adQ^a=
$$
$$
\int\Psi^*_k(Q,\,t)\,\Psi_{k'}(Q,\,t)\,
 \delta(\mu-f(Q)-k)\,\delta(\mu-f(Q)-k')\,M(\mu,\,Q)\,
 dk\,dk'\,d\mu\,\prod_adQ^a=
$$
\begin{equation}
\label{Psi_norm}
=\int\Psi^*_k(Q,\,t)\,\Psi_k(Q,\,t)\,
 M(f(Q)+k,\,Q)\,dk\,\prod_adQ^a=1.
\end{equation}
The solution (\ref{GS-A}) is normalizable under the condition that
$\Psi_k(Q,\,t)$ is a sufficiently narrow packet over $k$. Let us
emphasize that the dependence of $\Psi_k(Q,\,t)$ on $k$ is not
fixed by the equation (\ref{phys.SE}) in the sense that
$\Psi_k(Q,\,t)$ can be multiplied by an arbitrary function of $k$.
One cannot choose the function $\Psi_k(Q,\,t)$ to be not depending
on $k$, since in this case one would obtain a non-normalizable,
non-physical state. It would imply that the gauge condition is
fixed absolutely precisely ($\delta$-shaped packet), and such a
situation is unrealistic from the physical point of view. We should
rather consider a narrow enough packet over $k$ to fit a certain
classical $\bar k$ value:
\begin{eqnarray}
\Psi(\mu,Q,\theta,\bar\theta;\,t)&\!=\!&
 \frac1{\sqrt{2i\alpha\sqrt\pi}}\,
 \int\limits_{-\infty}^{\infty}
  \exp\left[-\frac1{2\alpha^2}\left(k-\bar k\right)^2\right]
  \Psi_{\bar k}(Q,\,t)\delta(\mu-f(Q)-k)
  \left(\bar\theta+i\theta\right)dk=\nonumber\\
\label{pack}
&\!=\!&\frac1{\sqrt{2i\alpha\sqrt\pi}}\,
 \exp\left[-\frac1{2\alpha^2}\left(\mu-f(Q)-\bar k\right)^2\right]\,
 \Psi_{\bar k}(Q,\,t)\,\left(\bar\theta+i\theta\right).
\end{eqnarray}

Since our investigation aims in giving description of a physical
Universe, we can introduce a density matrix
\begin{equation}
\label{rho_general}
\rho(Q,Q',t)=
 \int\Psi^*(\mu,\,Q,\,\theta,\,\bar\theta;\,t)\,
  \Psi(\mu,\,Q',\,\theta,\,\bar\theta;\,t)\,M(\mu,\,Q)\,
  d\mu\,d\theta\,d\bar\theta.
\end{equation}
For the wave function (\ref{pack}) the expression for density
matrix reads
\begin{eqnarray}
\rho(Q,Q',t)&\!=\!&
 \frac1{\alpha\sqrt\pi}\,
 \int\limits_{-\infty}^{\infty}
  \exp\left[-\frac1{2\alpha^2}\left(\mu-f(Q)-\bar k\right)^2
   -\frac1{2\alpha^2}\left(\mu-f(Q')-\bar k\right)^2\right]
 \times\nonumber\\
&\!\times\!&\Psi^*_{\bar k}(Q,\,t)\,\Psi_{\bar k}(Q',\,t)\,
 M(\mu,Q)\,d\mu=\nonumber\\
&\!=\!&\frac1{\alpha\sqrt\pi}\,
 \exp\left(-\frac1{4\alpha^2}\left[f(Q)-f(Q')\right]^2\right)\,
 \Psi^*_{\bar k}(Q,\,t)\,\Psi_{\bar k}(Q',\,t)\,\times\nonumber\\
&\!\times\!&\int\limits_{-\infty}^{\infty}
 \exp\left[-\frac1{\alpha^2}
  \left(\mu-\frac12\left[f(Q)-f(Q')\right]-\bar k\right)^2\right]\,
 M(\mu,Q)\,d\mu=\nonumber\\
&\!=\!&\exp\left(-\frac1{4\alpha^2}\left[f(Q)-f(Q')\right]^2\right)\,
 \Psi^*_{\bar k}(Q,\,t)\,\Psi_{\bar k}(Q',\,t)\times\nonumber\\
\label{rho_part}
&\!\times\!&M\left(\frac12\left[f(Q)+f(Q')\right]+\bar k,\,Q\right).
\end{eqnarray}
We have taken into account that only the vicinity of the ``point''
$\mu=\displaystyle\frac12\left[f(Q)+f(Q')\right]+\bar k$ gives a
significant contribution to the integral over $\mu$ and replaced
$\mu$ by its approximate value in the measure $M(\mu,Q)$. The
normalization condition for the density matrix is
\begin{equation}
\label{rho_norm}
\int\rho(Q,Q,t)\prod_adQ^a=
 \int\Psi^*_{\bar k}(Q,\,t)\,\Psi_{\bar k}(Q,\,t)\,
  M\left(f(Q)+\bar k,\,Q\right)\prod_adQ^a=1,
\end{equation}
it corresponds to the condition (\ref{Psi_norm}).

Thus, we can see that the density matrix contains the factor
\begin{equation}
\label{rho_factor}
\exp\left(-\frac1{4\alpha^2}\left[f(Q)-f(Q')\right]^2\right)
\end{equation}
so, if we choose $f(Q)$ to be equal to one of physical variables
which we shall denote as $q$ (it may be, for example, a scale
factor $a$), $f(Q)=q$, the expression (\ref{rho_factor}) will take
a Gaussian form
\begin{equation}
\label{Gauss_factor}
\rho\sim\exp\left[-\frac{(q-q')^2}{4\alpha^2}\right].
\end{equation}
its width $\sqrt 2\alpha$ being determined by a precision with
which we can fix the gauge condition. We shall not discuss here
what values $\alpha$ could take; it is a subject of quantum theory
of measurements.

At first glance, it may seem that the density matrix has a Gaussian
peak only in case of rather specific choice of a gauge condition
$\mu=q$. In fact, however, when the gauge condition (\ref{frame_A})
is fixed, it does not mean that some reference frame has been
already chosen. By the choice of a reference frame we imply,
following to Landau and Lifshitz, imposing some conditions on the
metric components $g_{0\mu}$ or, in our simplified model, on the
lapse function $N$. A reference frame is completely fixed only if
the choice of the parameterization function (\ref{paramet}), as
well as the gauge (\ref{frame_A}), is made. Let us emphasize again
that the choice of gauge variable parameterization and that of
gauge condition have an inseparable interpretation, -- they are
both determined by the construction of a clock, so without loss of
generality any gauge condition can be turned to $\mu=q$ by choosing
the function $v(\mu,Q)$. It is confirmed mathematically by the fact
that the Hamiltonian operator in physical subspace (\ref{phys.H-A})
after substitution $\mu=f(Q)+k$ depends on the function
$v\left(f(Q)+k,\,Q\right)$, but not on the functions $v(\mu,Q)$,
$f(Q)$ separately. Indeed, in (\ref{phys.H-A}) the quantum
correction $V$ can be presented in the form
\begin{eqnarray}
\left.V\right|_{\mu=f(Q)+k}&\!=\!&
 \frac5{12w^2}
  \frac{\delta w}{\delta Q_a}\frac{\delta w}{\delta Q^a}+
 \frac1{3w}\frac{\delta^2 w}{\delta Q_a\delta Q^a}+
 \frac{K-2}{6vw}
  \frac{\delta w}{\delta Q_a}\frac{\delta v}{\delta Q^a}-
  \nonumber\\
\label{phys.V}
&\!-\!&\frac{K^2-7K+6}{24v^2}
  \frac{\delta v}{\delta Q_a}\frac{\delta v}{\delta Q^a}+
 \frac{1-K}{6v}\frac{\delta^2 v}{\delta Q_a\delta Q^a},
\end{eqnarray}
where
$\displaystyle\frac{\delta v}{\delta Q^a}=
\left.\left(v_{,\mu}f_{,a}+v_{,a}\right)\right|_{\mu=f(Q)+k}$
is a total derivative with respect to $Q^a$ of the function
$v\left(f(Q)+k,\,Q\right)$, etc. Splitting the procedure of fixing
a reference frame into choosing the parameterization and imposing a
gauge condition is quite conventional, we shall give some examples
in the next section. However, the physical part of the wave
function $\Psi_k(Q,t)$, satisfying the equation (\ref{phys.SE}),
does not depend on this splitting, but on a chosen reference frame
as a whole.

The condition $\mu=q$ implies that the gauge subsystem interacts
with the physical subsystem through the variable $q$. As a result
of this interaction, the density matrix becomes about diagonal in
$q$. So, the Universe can be regarded as a classical system in any
region where the physical part of the wave function $\Psi_k(Q,t)$
can be represented in the WKB form with respect to the variable $q$
and where the gauge condition $\mu=q$ can be imposed. At the same
time, the density matrix does not diagonalize in other physical
variables (such as gravitational waves and matter fields, with $q$
representing the scale factor), which do not interact with the
gauge subsystem and should be treated quantum mechanically. One can
assume that these degrees of freedom describe small perturbations
on the background of an isotropic universe. A similar approach is
adopted, for example, in \cite{Halliwell} where the long-wavelength
modes of a scalar field play the role of environment while the
short-wavelength modes remain quantum mechanical. The authors of
the works \cite{Halliwell, FM} and others seek for a model in which
a density matrix would contain a Gaussian factor. However, it is
always questionable what part of the physical Universe should be
considered as an environment. On the other side, taking into
account the gauge subsystem is thought to be inevitable when
constructing mathematically consistent quantum geometrodynamics.
Gauge degrees of freedom are not observable directly, but only by
its influence on the physical subsystem. The result
(\ref{Gauss_factor}), namely, that the density matrix may have a
Gaussian peak in some variable under physically reasonable fixing
of gauge condition, is quite general. It seems, therefore, to be
natural to regard gauge degrees of freedom as the environment for
the physical Universe. One could say, in a certain sense, that the
physical Universe is a subsystem of itself.

\section{The gauge subsystem as a factor of cosmological evolution}
When deriving the Schr\"odinger equation from the path integral
with the effective action (\ref{action}), we approximate the path
integral on extended gauged set of equations obtained by varying
this effective action. The extended set of equations includes
ghosts equations and a gauge condition, and equations for physical
degrees of freedom also contain gauge-noninvariant terms. So, the
gauged Einstein equations look like
\begin{equation}
\label{Ein.eqs}
R_{\mu}^{\nu}-\frac12\delta_{\mu}^{\nu}R=
 \kappa\left(T_{\mu(mat)}^{\nu}+T_{\mu(obs)}^{\nu}+T_{\mu(ghost)}^{\nu}\right),
\end{equation}
where $T_{\mu(mat)}^{\nu}$ is the energy-momentum tensor of matter
fields, $T_{\mu(obs)}^{\nu}$ and $T_{\mu(ghost)}^{\nu}$ are
obtained by varying the gauge-fixing and ghost action,
respectively. $T_{\mu(obs)}^{\nu}$ describes the observer (the
gauge subsystem) in the extended set of equations.

In particular, the $0\choose 0$-Einstein equation (Hamiltonian
constraint) is transformed to the form $H=E$, where $H$ is a
Hamiltonian in extended phase space and
\begin{equation}
\label{eigenvalue}
E=-\int\sqrt{-g}\,T_{0(obs)}^0\,d^3 x.
\end{equation}
In quantum theory the modified Hamiltonian constraint leads to a
stationary Schr\"odinger equation (see (\ref{stat.states})).

To give a simple example, in this section we shall bear in mind an
isotropic universe, then the parameterization function $v(\mu,Q)$,
as well as the gauge-fixing function $f(Q)$, will depend only on a
scale factor, i. e.
\begin{equation}
\label{isotr.paramet}
\displaystyle\frac{a^3}{N}=v(\mu,a),\quad
\mu=f(a)+k.
\end{equation}
The quasi-energy-momentum tensor of the gauge subsystem reads:
\begin{equation}
\label{T_obs}
T_{\mu(obs)}^{\nu}=
 {\rm diag}\left(\varepsilon_{(obs)},\,
  -p_{(obs)},\,-p_{(obs)},\,-p_{(obs)}\right);
\end{equation}
\begin{equation}
\label{eps_obs}
\varepsilon_{(obs)}=
 -\left.\frac{\dot\pi_0}{2\pi^2}
  \frac{v^2(\mu,a)}{a^6 v_{,\mu}}\right|_{\mu=f(a)+k};
\end{equation}
\begin{equation}
\label{p_obs}
p_{(obs)}=\varepsilon_{(obs)}
 \left.\left[1-\frac{a}{3v(\mu,a)}
  \left(v_{,\mu}f_{,a}+v_{,a}\right)\right]\right|_{\mu=f(a)+k}.
\end{equation}
The last formula gives the equation of state for the gauge
subsystem depending on parameterization and gauge condition. Note
that, again, the equation of state after substitution $\mu=f(a)+k$
depends on the function $v\left(f(a)+k,\,a\right)$, but not on the
functions $v(\mu,a)$, $f(a)$ separately.

Let us choose
\begin{equation}
\label{choice1}
v(\mu,a)=\frac{a^2}{\mu},\quad
\mu=1+\frac1{a^4}.
\end{equation}
(Here and below we shall assume that the classical value
$\bar k=0$). As follows from (\ref{isotr.paramet}), it corresponds
to the condition for the lapse function $N$:
\begin{equation}
\label{N-cond}
N=a+\frac1{a^3}.
\end{equation}
This gauge condition is rather interesting in some respects. For
large $a$ we have $N=a$ (conformal time gauge), while in the limit
of small $a$ the condition (\ref{N-cond}) can be rewritten as
$N a^3=1$. The latter corresponds to the constraint on metric
components $\sqrt{-g}={\rm const}$. This constraint is known to
lead to the appearance of $\Lambda$-term in the Einstein equations
\cite{Weinberg, Shest2}. For $v(\mu,a)$ and $f(a)$ determined by
(\ref{choice1}) the equation of state (\ref{p_obs}) reduces to
\begin{equation}
\label{state.eq}
p_{(obs)}=\varepsilon_{(obs)}
 \left[1-\frac13\left(\frac4{1+a^4}+2\right)\right].
\end{equation}
In the course of cosmological evolution the equation of state
changes from $p_{(obs)}=-\varepsilon_{(obs)}$ in the limit of small
$a$ to $p_{(obs)}=\displaystyle\frac13\varepsilon_{(obs)}$ in the
limit of large $a$. The former corresponds a medium with negative
pressure typical for an exponentially expanded early universe with
$\Lambda$-term, the latter is an ultrarelativistic equation of
state, and the Einstein equations in the limit of large $a$ have a
solution describing a Friedmann universe in the conformal time
gauge $N=a$. Therefore, we can see that the gauge subsystem appears
to be a factor of cosmological evolution; its state changing over
the history of the Universe determining a cosmological scenario.

\section{Discussion}
If we adopt the ADM parameterization, namely, regard the lapse
function $N$ as a gauge variable and impose the condition
(\ref{N-cond}), according the above consideration, we shall come to
the conclusion that at large $a$, when $N=a$ in a good
approximation, the density matrix will have a Gaussian peak
(\ref{Gauss_factor}) with $q$ representing $a$. In other words, this
simple model confirms that the scale factor becomes a classical
variable in the region of large $a$.

On the other hand, we are free to choose another gauge variable,
$\mu$, giving the function $v(\mu,Q)$. Then, we shall change
properties of environment and the character of its interaction with
the physical subsystem. Remind that the density matrix
(\ref{rho_general}) is obtained by integrating out a gauge variable
defined by (\ref{paramet}). In particular, as was said above, for a
given reference frame one can choose $\mu$ to satisfy the condition
$\mu=a$. So, instead of (\ref{choice1}) one can put
\begin{equation}
\label{choice2}
v(\mu,a)=\frac{a^2}{1+\displaystyle\frac1{\mu^4}},\quad
\mu=a.
\end{equation}
It leads to the same condition (\ref{N-cond}) for $N$ and the same
equation of state (\ref{state.eq}). The density matrix will have a
Gaussian peak (\ref{Gauss_factor}) in the whole region where the
condition $\mu=a$ can be imposed, or, where the reference frame
determined by (\ref{N-cond}) can be chosen. (One could make the
simplest choice $N=a$, but in this case in the quasiclassical limit
one would get a Friedmann universe without $\Lambda$-term and,
correspondingly, without a stage of inflation.) The choice
(\ref{choice2}) means that we choose in (\ref{GS-A}) the basis,
which is the set of eigenstates of the gauge operator $(\mu-a)$:
\begin{equation}
\label{frame_B}
\left(\mu-a\right)|k\rangle=k\,|k\rangle;\quad
|k\rangle=\delta\left(\mu-a-k\right).
\end{equation}
The basis (\ref{frame_B}) plays the role of a preferred basis in
the sense that the density matrix is about diagonal. In this case
we choose an environment causing the density matrix to diagonalize.
However, the variable $\mu$, which represents the environment, may
not have a clear physical meaning like the lapse function $N$.

The gauge subsystem has also other features, which are usually
implied for environment. In particular, in any region where some
gauge condition is fixed, a state of the Universe is not disturbed
by interaction with the gauge system. Indeed, since one of
canonical equations in extended phase space is a gauge condition in
a differential form (\ref{diff_form}), a gauge operator commutes
with the Hamiltonian (\ref{H}) \cite{SSV4}. For example, when the
state of the Universe is a specific state discussed above with
$E=0$, described by a quasiclassical wave function $\Psi_{k0}(Q)$,
it will not be disturbed by interaction with the environment.

To summarize, under a suitable definition of the parameterization
function $v(\mu,a)$, one can get that the density matrix would have
a Gaussian peak in some variable in any region where a certain
reference frame is chosen. An appropriate gauge condition may be
rather complicated, as the condition (\ref{N-cond}), which includes
different regimes in limiting cases. (In fact, Eq.(\ref{N-cond})
describes a transition from the condition $N a^3=1$ to $N=a$). In
simple cosmological models it is possible to introduce a single
reference frame in the whole spacetime, and the behavior of the
Universe can be regarded as classical almost over the whole its
history. The situation is different if one admits nontrivial
topology of spacetime. In general, spacetime manifold may consist
of regions covered by different coordinate charts, so that one
should introduce different reference frames in these regions. A
transition from one reference frame to another cannot be described
by a single gauge condition like (\ref{N-cond}), and the Universe
cannot behave in a classical manner nearby borders of regions where
different reference frames are introduced. It especially concerns
spacetime manifolds with horizons. Let us note that the
quasiclassical approximation is not valid nearby the borders of
these regions as well, so that the two requirements -- the
quasiclassical character of a wave function and fixing a certain
reference frame -- completely coincide.

In \cite{Shest3} we have considered a small variation of the
gauge-fixing function $f(Q)$ while the parameterization function
$v(\mu,Q)$ being fixed. This variation corresponds to a transition
to another reference frame and another basis
\begin{equation}
\label{frame_C}
\left(\mu-f(Q)-\delta f(Q)\right)|k\rangle=k\,|k\rangle;\quad
|k\rangle=\delta\left(\mu-f(Q)-\delta f(Q)-k\right).
\end{equation}
Then the Hamiltonian in physical subspace (\ref{phys.H-A}) acquires
additional terms, which contain anti-Hermitian part in the original
subspace defined by the basis (\ref{k-vector}). Accordingly, a
measure in the physical subspace $M\left(f(Q)+k,\,Q\right)$ (see
(\ref{Psi_norm})) changes to $M\left(f(Q)+\delta
f(Q)+k,\,Q\right)$. The change of the measure and the appearance of
an anti-Hermitian part of the Hamiltonian (\ref{phys.H-A}) show
that a transition to another reference frame has an irreversible
character. As a consequence of interaction with the subsystem of
gauge degrees of freedom, the Universe may not stay in one of
states of the superposition (\ref{stat.WF}), in particular, in a
special state with $E=0$ for which a classical limit could be
obtain. Its state would change in some irreversible way.

\end{document}